\documentclass{article}
\usepackage{waspaa23,amssymb,amsmath,url,times}
\usepackage{color}

\usepackage[dvipdfmx]{graphicx}
\usepackage{bm,cite,subfig}
\usepackage{algorithm,algorithmic}

\newcommand{\argmin}{\mathop{\rm arg~min}\limits}

\newcommand{\minimize}{\mathop{\rm minimize}\limits}

\title{Perceptual Quality Enhancement of Sound Field Synthesis Based on Combination of Pressure and Amplitude Matching}

\name{Keisuke Kimura,$^{1}$
      Shoichi Koyama,$^{2}$
      Hiroshi Saruwatari$^{1}$}
\address{$^1$ The University of Tokyo, 7-3-1 Hongo, Bunkyo-ku, Tokyo 113-8656, Japan\\              
         $^2$ National Institute of Informatics, 2-1-2 Hitotsubashi, Chiyoda-ku, Tokyo 101-8430, Japan\\
          koyama.shoichi@ieee.org\\
}

\begin{document}

\ninept
\maketitle

\begin{sloppy}

\begin{abstract}
 A sound field synthesis method enhancing perceptual quality is proposed. Sound field synthesis using multiple loudspeakers enables spatial audio reproduction with a broad listening area; however, synthesis errors at high frequencies called spatial aliasing artifacts are unavoidable. To minimize these artifacts, we propose a method based on the combination of pressure and amplitude matching. On the basis of the human's auditory properties, synthesizing the amplitude distribution will be sufficient for horizontal sound localization. Furthermore, a flat amplitude response should be synthesized as much as possible to avoid coloration. Therefore, we apply amplitude matching, which is a method to synthesize the desired amplitude distribution with arbitrary phase distribution, for high frequencies and conventional pressure matching for low frequencies. Experimental results of numerical simulations and listening tests using a practical system indicated that the perceptual quality of the sound field synthesized by the proposed method was improved from that synthesized by pressure matching. 
\end{abstract}


\section{Introduction}
\label{sec:intro}

Sound field synthesis/reproduction is a technique to recreate a spatial sound using multiple loudspeakers (or secondary sources). One of its prospective applications is spatial audio for virtual/augmented reality, which enables spatial audio reproduction with a broader listening area than in the case of conventional spatial audio techniques, such as multichannel surround sound and binaural synthesis. 

Sound field synthesis methods based on the minimization of the squared error between synthesized and desired sound fields, such as \textit{pressure matching} and \textit{mode matching}~\cite{Nelson:J_SV_1993,Kirkeby:JASA_J_1996,Betlehem:JASA_J_2005,Poletti:J_AES_2005,Ueno:IEEE_ACM_J_ASLP2019,Koyama:JAES2023}, have practical advantages compared with methods based on analytical representations derived from boundary integral equations such as \textit{wave field synthesis} and \textit{higher-order ambisonics}~\cite{Berkhout:JASA_J_1993,Spors:AES124conv,Daniel:AES114conv,Poletti:J_AES_2005,Ahrens:Acustica2008,Wu:IEEE_J_ASLP2009,Koyama:IEEE_J_ASLP2013,Koyama:IEEE_ACM_J_ASLP2014}, since the array geometry of the secondary sources can be arbitrary. In particular, pressure matching is widely used because of its simple implementation. 

A well-known issue of the sound field synthesis methods is \textit{spatial aliasing artifacts}. That is, depending on the secondary source placement, the synthesis accuracy can significantly decrease at high frequencies, which can lead to the degradation of sound localization for listeners and distortion of timbre, or \textit{coloration}, of source signals. Thus, the perceptual quality of the synthesized sound field can considerably deteriorate. 

To improve the perceptual quality, we propose a method combining \textit{amplitude matching}, which was originally proposed for multizone sound field control~\cite{Koyama:ICASSP2021,Abe:IEEE_ACM_J_ASLP2023}, with pressure matching. Amplitude matching is a method to synthesize the desired amplitude (or magnitude) distribution, leaving the phase distribution arbitrary, whereas pressure matching aims to synthesize both amplitude and phase distributions, i.e., pressure distribution. We apply amplitude matching to mitigate the spatial aliasing artifacts by reducing the parameters to be controlled at high frequencies, keeping the range of the listening area broad. On the basis of the human's auditory properties, the interaural level difference (ILD) is known to be a dominant cue for horizontal sound localization at high frequencies, compared with the interaural time difference (ITD)~\cite{Blauert:SpatialHearing,Brughera:JASA2013}. Therefore, synthesizing the amplitude distribution will be sufficient for sound localization. Furthermore, by prioritizing the synthesis of the desired amplitude distribution, a flat amplitude response is reproduced as much as possible, and coloration effects can be alleviated. We formulate a new cost function combining amplitude matching for high frequencies and conventional pressure matching for low frequencies, which can be solved in a similar manner to amplitude matching. We evaluate the proposed method through numerical experiments in the frequency domain and listening experiments in a real environment. 

\section{Sound Field Synthesis Problem}
\label{sec:problem}

Let $\Omega\subset\mathbb{R}^3$ be a target region for synthesizing the desired sound field. As shown in Fig.~\ref{fig:sfs}, $L$ secondary sources are placed at $\bm{r}_l \in \mathbb{R}^3\backslash\Omega$ ($l\in\{1,\ldots,L\}$). The driving signal of the $l$th secondary source and its transfer function to the position $\bm{r}\in\Omega$ at the angular frequency $\omega$ are denoted as $d_l(\omega)$ and $g_l(\bm{r},\omega)$, respectively. Then, the synthesized pressure distribution $u_{\mathrm{syn}}(\bm{r},\omega)$ ($\bm{r}\in\Omega$) is represented as
\begin{align}
 u_{\mathrm{syn}}(\bm{r},\omega) &= \sum_{l=1}^L d_l(\omega) g_l(\bm{r},\omega) \notag\\
&= \bm{g}(\bm{r},\omega)^{\mathsf{T}} \bm{d}(\omega),
\end{align}
where $\bm{g}(\bm{r},\omega)\in\mathbb{C}^L$ and $\bm{d}(\omega)\in\mathbb{C}^L$ are the vectors consisting of $g_l(\bm{r},\omega)$ and $d_l(\omega)$, respectively. Hereafter, $\omega$ is omitted for notational simplicity. 

The objective is to synthesize the desired sound field $u_{\mathrm{des}}(\bm{r})$ over the target region $\Omega$. We define the optimization problem to obtain $\bm{d}$ as follows.
\begin{align}
 \minimize_{\bm{d}\in\mathbb{C}^L} Q(\bm{d}) := \int_{\bm{r}\in\Omega} \left| \bm{g}(\bm{r})^{\mathsf{T}}\bm{d} - u_{\mathrm{des}}(\bm{r}) \right|^2 \mathrm{d}\bm{r}
\label{eq:cost_synth}
\end{align}
Since this problem is difficult to solve owing to the regional integration, several methods to approximately solve it, for example, (weighted) pressure and mode matching~\cite{Nelson:J_SV_1993,Kirkeby:JASA_J_1996,Betlehem:JASA_J_2005,Poletti:J_AES_2005,Ueno:IEEE_ACM_J_ASLP2019,Koyama:JAES2023}, have been proposed. 

\begin{figure}[t]
  \centering
  \includegraphics[width=130pt,clip]{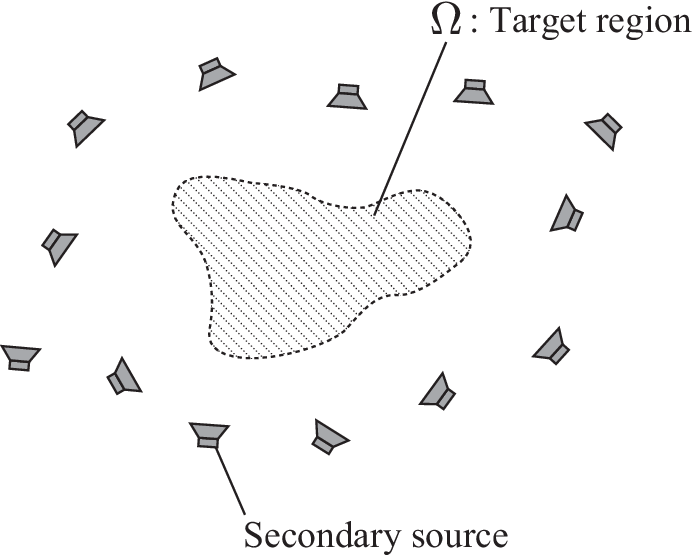}
  \caption{Sound field synthesis over the target region $\Omega$ with multiple loudspeakers}
  \label{fig:sfs}
  \vspace{-10pt}
\end{figure}

\subsection{Pressure Matching}
\label{sec:PM/AM}
\vspace{-4pt}

Pressure matching is one of the widely used sound field synthesis methods to approximately solve \eqref{eq:cost_synth}. First, the region $\Omega$ is discretized into $N$ control points whose positions are denoted as $\{\bm{r}_n\}_{n=1}^N$. We assume that the control points are arranged densely enough over $\Omega$. Then, the cost function for pressure matching is formulated as the minimization problem of the squared error between the synthesized and desired pressures at the control points as
\begin{align}
 \minimize_{\bm{d}\in\mathbb{C}^L}\ \|\bm{G}\bm{d}-\bm{u}_{\mathrm{des}}\|_2^2+\beta\|\bm{d}\|_2^2,
\label{eq:cost_pm}
\end{align}
where $\bm{G}\in\mathbb{C}^{N\times L}$ is the matrix consisting of transfer functions $\{g_l(\bm{r}_n)\}_{n=1}^N$, $\bm{u}_{\mathrm{des}}\in\mathbb{C}^N$ is the vector consisting of the desired pressures $\{u_{\mathrm{des}}(\bm{r}_n)\}_{n=1}^N$, and $\beta$ is the regularization parameter. This least-squares problem \eqref{eq:cost_pm} has a closed-form solution as follows:
\begin{align}
\hat{\bm{d}}=\left(\bm{G}^\mathsf{H}\bm{G}+\beta\bm{I}\right)^{-1}\bm{G}^\mathsf{H}\bm{u}_{\mathrm{des}}.
\label{eq:d_PM}
\end{align}
Pressure matching is extended as weighted pressure matching~\cite{Koyama:ICA2022,Koyama:JAES2023} by combining with sound field interpolation techniques~\cite{Ueno:IEEE_SPL2018,Ueno:IEEE_J_SP2021}. 

Another strategy to approximate $Q(\bm{d})$ is to represent the sound field by spherical wavefunction expansion~\cite{Williams:FourierAcoust,Martin:MultScat}, which is referred to as (weighted) mode matching~\cite{Poletti:J_AES_2005,Ueno:IEEE_ACM_J_ASLP2019}. 
It is demonstrated that weighted pressure matching is a special case of weighted mode matching in \cite{Koyama:JAES2023}.

\subsection{Spatial Aliasing Artifacts}
\label{sec:aliasing}
\vspace{-4pt}

On the basis of the single-layer potential~\cite{Colton:InvAcoust_2013,Williams:FourierAcoust}, the desired sound field can be perfectly synthesized if secondary sources are continuously distributed point sources on a surface of the target region $\Omega$. However, owing to the discrete placement of the secondary sources, spatial aliasing artifacts are unavoidable for sound field synthesis methods. The synthesis accuracy can decrease particularly at high frequencies, which can lead to the degradation of sound localization and the coloration of source signals. The properties of spatial aliasing in analytical sound field synthesis methods have been extensively investigated~\cite{Spors:AES2006conv}, but spatial aliasing can also occur in numerical methods, depending on the secondary source placement~\cite{Koyama:IEEE_ACM_J_ASLP2020,Kimura:WASPAA2021}. One of the strategies to improve the synthesis accuracy at high frequencies is to make the target region smaller~\cite{Poletti:Internoise2014,Ueno:IEEE_ACM_J_ASLP2019}. The challenge here is to improve the perceptual quality of the synthesized sound field by mitigating spatial aliasing artifacts in a broad listening area. 

\section{Proposed Method Based on Combination of Pressure and Amplitude Matching}
\label{sec:proposed}
\vspace{-6pt}

Even when it is difficult to synthesize the desired pressure distribution, i.e., amplitude and phase, for high frequencies, it can be considered that synthesizing only the amplitude distribution can be achieved by leaving the phase distribution arbitrary. On the basis of the human's auditory properties, ILD is known to be a dominant cue for horizontal sound localization above $1500~\mathrm{Hz}$, and the dependence on ITD is markedly reduced from $1000$--$1500~\mathrm{Hz}$~\cite{Blauert:SpatialHearing,Brughera:JASA2013}, which indicates that synthesizing the amplitude distribution is sufficient for sound localization above $1500~\mathrm{Hz}$. 
This perceptual property is also used in a method for binaural rendering~\cite{Schorkhuber:DAGA2018}.
Furthermore, coloration effects can be alleviated by reproducing the flat amplitude response as much as possible. Therefore, we propose a method combining amplitude matching~\cite{Koyama:ICASSP2021,Abe:IEEE_ACM_J_ASLP2023}, which is aimed to synthesize the desired amplitude at the control points, with pressure matching. 
By applying amplitude matching for high frequencies, the parameters to be controlled can be reduced from pressure matching, keeping the range of the target region, i.e., the number of control points. Thus, we can improve the perceptual quality of the synthesized sound field by reproducing a more accurate amplitude distribution over $\Omega$.

\vspace{-4pt}
\subsection{Proposed Algorithm}
\label{sec:proposed-alg}
\vspace{-4pt}

We define the optimization problem of the proposed method as a composite of pressure and amplitude matching as
\begin{align}
    \minimize_{\bm{d}\in\mathbb{C}^L} J(\bm{d}) := &\left(1-\gamma\right) \left\| \bm{G}\bm{d}-\bm{u}_{\mathrm{des}} \right\|_2^2 \notag\\ 
& \ \ + \gamma \left\| |\bm{G}\bm{d}| - |\bm{u}_{\mathrm{des}}| \right\|_2^2 + \beta\|\bm{d}\|_2^2, 
\label{eq:cost_pm+am}
\end{align}
where $|\cdot|$ represents the element-wise absolute value of vectors, and $\gamma \in \mathbb{R}([0,1])$ is the parameter that determines the balance between pressure and amplitude matching. When $\gamma=0$, \eqref{eq:cost_pm+am} corresponds to pressure matching, and when $\gamma=1$, it corresponds to amplitude matching; therefore, $\gamma$ should be set close to $0$ for low frequencies and $1$ for high frequencies. For example, $\gamma$ can be defined as a sigmoid function of $\omega$ with the transition angular frequency $\omega_{\mathrm{T}}$ and parameter $\sigma$ as
\begin{align}
    \gamma(\omega)=\frac{1}{1+\mathrm{e}^{-\frac{\sigma}{2\pi}(\omega-\omega_{\mathrm{T}})}}.
\label{eq:freq_trans}
\end{align}

Since the cost function $J$ in \eqref{eq:cost_pm+am} is neither convex nor differentiable owing to the squared error term of the amplitude matching, \eqref{eq:cost_pm+am} does not have a closed-form solution. However, the alternating direction method of multipliers (ADMM) can be applied in a similar manner to the algorithm for amplitude matching~\cite{Abe:IEEE_ACM_J_ASLP2023}. First, we introduce the auxiliary variables of amplitude $\bm{a}\in\mathbb{R}_{\geq\bm{0}}^N$ and phase $\bm{\theta}\in\mathbb{R}^N$ such that $\bm{G}\bm{d}=\bm{a}\odot \mathrm{e}^{{\mathrm{j}}\bm{\theta}}$, where $\odot$ represents the Hadamard product. Then, \eqref{eq:cost_pm+am} is reformulated as
\begin{align}
    &\minimize_{\bm{d},\bm{a},\bm{\theta}} \ (1-\gamma)\left\|\bm{a} \odot \mathrm{e}^{{\mathrm{j}} \bm{\theta}}-\bm{u}_{\mathrm{des}}\right\|_2^2 \notag\\ 
&\hspace{60pt}+\gamma\left\|\bm{a}-|\bm{u}_{\mathrm{des}}|\right\|_2^2 + \beta\|\bm{d}\|_2^2 \notag\\
&\mathrm{subject \ to} \ \bm{G}\bm{d}=\bm{a} \odot \mathrm{e}^{{\mathrm{j}}\bm{\theta}}.
\label{eq:opt_admm}
\end{align}
The augmented Lagrangian function $L_{\rho}$ for \eqref{eq:opt_admm} is defined as
\begin{align}
    &L_\rho(\bm{a}, \bm{\theta}, \bm{d}, \bm{\lambda}) \notag\\
    &= (1-\gamma)\left\|\bm{a} \odot \mathrm{e}^{\mathrm{j} \bm{\theta}} - \bm{u}_{\mathrm{des}} \right\|_2^2 + \gamma \left\| \bm{a} - |\bm{u}_{\mathrm{des}}| \right\|_2^2 + \beta\|\bm{d}\|_2^2 \notag\\
    &\hspace{20pt} + \Re\left[ \bm{\lambda}^\mathsf{T}(\bm{G}\bm{d}-\bm{a} \odot \mathrm{e}^{\mathrm{j}\bm{\theta}}) \right] + \frac{\rho}{2} \|\bm{G}\bm{d}-\bm{a} \odot \mathrm{e}^{\mathrm{j}\bm{\theta}}\|_2^2,
\label{eq:aug_lagrange}
\end{align}
where $\bm{\lambda}\in\mathbb{C}^N$ is the Lagrange multiplier, $\Re[\cdot]$ represents the real part of a complex value, and $\rho>0$ is the penalty parameter. 

Each variable is alternately updated on the basis of ADMM as
\begin{align}
    &\left(\bm{a}^{(i+1)},\bm{\theta}^{(i+1)}\right) = \argmin_{\bm{a},\bm{\theta}} \ L_\rho\left(\bm{a},\bm{\theta},\bm{d}^{(i)},\bm{\lambda}^{(i)}\right) \label{eq:ADMM_u_theta}\\
    &\bm{d}^{(i+1)} = \argmin_{\bm{d}} \ L_\rho\left(\bm{a}^{(i+1)},\bm{\theta}^{(i+1)},\bm{d},\bm{\lambda}^{(i)}\right) \label{eq:ADMM_d}\\
    &\bm{\lambda}^{(i+1)} = \bm{\lambda}^{(i)} + \rho\left(\bm{G}\bm{d}^{(i+1)}-\bm{a}^{(i+1)} \odot \mathrm{e}^{\mathrm{j}\bm{\theta}^{(i+1)}}\right), \label{eq:ADMM_lambda}
\end{align}
where $i$ denotes the iteration index. \eqref{eq:ADMM_u_theta} is minimized independently for $\bm{\theta}$ and $\bm{a}$ as 
\begin{align}
 &\bm{\theta}^{(i+1)} =\arg \left((1-\gamma)\bm{u}_{\mathrm{des}} + \frac{\rho}{2} \left(\bm{G}\bm{d}^{(i)} + \frac{\bm{\lambda}^{(i)}}{\rho} \right)\right), \label{eq:theta} \\
 &\bm{a}^{(i+1)}
    = \frac{\left|2(1-\gamma)\bm{u}_{\mathrm{des}} + \rho\left(\bm{G}\bm{d}^{(i)}+\frac{\bm{\lambda}^{(i)}}{\rho}\right)\right| + 2\gamma|\bm{u}_{\mathrm{des}}|}{\rho+2}. \label{eq:u}
\end{align}
The update rule for $\bm{d}$ is obtained by solving \eqref{eq:ADMM_d} as
\begin{align}
 &\bm{d}^{(i+1)}\notag\\
 & \ \ = \left(\bm{G}^\mathsf{H}\bm{G} + \frac{2\beta}{\rho}\bm{I}\right)^{-1} \bm{G}^\mathsf{H} \left(\bm{a}^{(i+1)} \odot \mathrm{e}^{{\mathrm{j}} \bm{\theta}^{(i+1)}} - \frac{\bm{\lambda}^{(i)}}{\rho}\right). \label{eq:d}
\end{align}
By iteratively updating $\bm{\theta}^{(i)}$, $\bm{a}^{(i)}$, $\bm{d}^{(i)}$, and $\bm{\lambda}^{(i)}$ by using \eqref{eq:theta}, \eqref{eq:u}, \eqref{eq:d}, and \eqref{eq:ADMM_lambda}, respectively, starting with initial values, we can obtain the optimal driving signal $\bm{d}$.

\subsection{Time-Domain Filter Design}

By using the proposed algorithm described in Sect.~\ref{sec:proposed-alg}, we can obtain the driving signal in the frequency domain. In practice, a finite impulse response (FIR) filter is obtained by computing the inverse Fourier transform of $\bm{d}$ for target frequency bins. However, if the driving signal is independently determined for each frequency, it can have discontinuities between frequency bins particularly for $\gamma=1$ because the phase of $\bm{d}$ is arbitrary. These discontinuities can lead to an unnecessarily large FIR filter length. 

To overcome this issue, the differential-norm penalty, which is also used in amplitude matching~\cite{Abe:IEEE_ACM_J_ASLP2023}, can be applied. By introducing the subscript of the index of the frequency bin $f\in\{1,\ldots,F\}$, we define the differential-norm penalty for the $f$th frequency bin as
\begin{align}
 D(\bm{d}_f) = \| \bm{d}_f - \bm{d}_{f-1}\|_2^2.
\end{align}
The optimization problem for the time-domain filter design is represented as
\begin{align}
    &\minimize_{ \{\bm{d}_f\}_{f=1}^F}
    \sum_{f=1}^F \left[ (1-\gamma)\left\|\bm{G}_{f}\bm{d}_f-\bm{u}_{{\mathrm{des}},f}\right\|_2^2 \right.\notag\\
    & \hspace{10pt} \left. + \gamma\left\||\bm{G}_{f}\bm{d}_f|-|\bm{u}_{{\mathrm{des}},f}|\right\|_2^2 \right] + \alpha \sum_{f=2}^F D(\bm{d}_f) + \beta \sum_{f=1}^F \|\bm{d}_f\|_2^2,
\label{eq:costfunc_diff}
\end{align}
where $\alpha$ is the weight for the differential-norm penalty term. ADMM is similarly applied to solve \eqref{eq:costfunc_diff}, but its detailed derivation is omitted. The derivation of ADMM for amplitude matching and a technique to reduce computational complexity can be found in~\cite{Abe:IEEE_ACM_J_ASLP2023}.

\section{Experiments}

We conducted experiments to evaluate the proposed method (Proposed) compared with pressure matching (PM). First, numerical experimental results are shown to evaluate the ILD of a listener and the amplitude response of the synthesized sound field. Second, the results of listening experiments using a practical system are presented. 

\subsection{Numerical Experiments}

\begin{figure}[t]
  \centering
  \includegraphics[width=140pt]{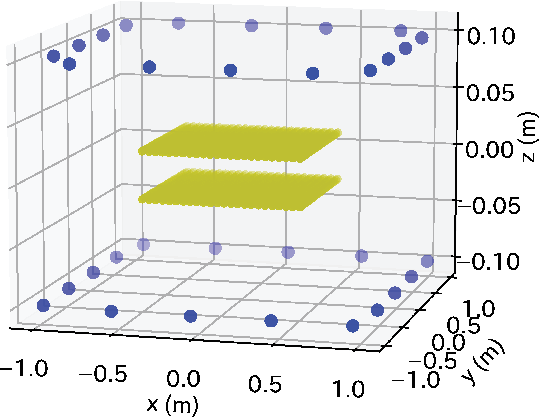}
  \caption{Experimental setup. Blue circles and yellow dots represent sources and control points, respectively.}
  \label{fig:field3D}
\end{figure}

Numerical experiments in the frequency domain were conducted under the three-dimensional free-field assumption. The target region $\Omega$ was a cuboid of $1.0~\mathrm{m} \times 1.0~\mathrm{m} \times 0.04~\mathrm{m}$. As shown in Fig.~\ref{fig:field3D}, 16 omnidirectional loudspeakers were placed along the borders of the squares of $2.0~\mathrm{m} \times 2.0~\mathrm{m}$ at the heights of $z=\pm 0.1~\mathrm{m}$, and $24 \times 24 \times 2$ control points were regularly placed inside $\Omega$. Therefore, the total number of loudspeakers and control points were $L=32$ and $N=1152$, respectively. The desired sound field was a spherical wave from the point source at $(2.0, 0.0, 0.0)~\mathrm{m}$. In Proposed, $\gamma$ was set using \eqref{eq:freq_trans} with $\omega_{\mathrm{T}}/2\pi=2000$ and $\sigma=0.01$ so that the phase distribution becomes arbitrary above $2000~\mathrm{Hz}$. The regularization parameter $\beta$ for Proposed and PM was set as $\|\bm{G}^{\mathsf{H}}  \bm{G}\|_2^2 \times10^{-3}$. The penalty parameter $\rho$ in \eqref{eq:aug_lagrange} was $1.0$.

We evaluated the ILDs of the synthesized sound field  when a listener's head was in $\Omega$. The binaural signals at $\omega$ for the position $\bm{r}_{\mathrm{H}}\in\Omega$ and azimuth direction $\phi_{\mathrm{H}} \in [0,2\pi)$ of the listener's head were denoted as $b_{\mathrm{L}}(\bm{r}_{\mathrm{H}},\phi_{\mathrm{H}},\omega)$ and $b_{\mathrm{R}}(\bm{r}_{\mathrm{H}},\phi_{\mathrm{H}},\omega)$ for the left and right ears, respectively. $b_{\mathrm{L}}$ and $b_{\mathrm{R}}$ were obtained by calculating the transfer function between the loudspeakers and the listener's ears using Mesh2HRTF~\cite{ziegelwanger2015mesh2hrtf,ziegelwanger2015numerical}. The ILD for $\bm{r}_{\mathrm{H}}$ and $\phi_{\mathrm{H}}$ was calculated in the frequency domain as
\begin{align}
    \mathrm{ILD}(\bm{r}_{\mathrm{H}},\phi_{\mathrm{H}}) &:= 10\log_{10} \frac{\sum_{\omega}|b_\mathrm{L}(\bm{r}_{\mathrm{H}},\phi_{\mathrm{H}},\omega)|^2}{\sum_{\omega}|b_\mathrm{R}(\bm{r}_{\mathrm{H}},\phi_{\mathrm{H}},\omega)|^2}.
\label{eq:MSE_def}
\end{align}
The evaluation measure was the normalized error ($\mathrm{NE}$) between the synthesized and desired ILDs ($\mathrm{ILD}_{\mathrm{syn}}$ and $\mathrm{ILD}_{\mathrm{true}}$, respectively) at $\bm{r}_{\mathrm{H}}$ defined as
\begin{align}
    \mathrm{NE}(\bm{r}_{\mathrm{H}}) = \frac{\sum_{\phi_{\mathrm{H}}}|\mathrm{ILD}_{\mathrm{syn}}(\bm{r}_{\mathrm{H}},\phi_{\mathrm{H}}) - \mathrm{ILD}_{\mathrm{true}}(\bm{r}_{\mathrm{H}},\phi_{\mathrm{H}})|}{\sum_{\phi_{\mathrm{H}}}|\mathrm{ILD}_{\mathrm{true}}(\bm{r}_{\mathrm{H}},\phi_{\mathrm{H}})|},
\end{align}
where the summation for $\phi_{\mathrm{H}}$ was calculated for $\{0,\pi/32,\pi/16,\ldots,31\pi/32\}~\mathrm{rad}$. The distributions of $\mathrm{NE}(\bm{r}_{\mathrm{H}})$ on the $x$--$y$-plane at $z=0$ for Proposed and PM are shown in Fig.~\ref{fig:ILD_NE}. The evaluated positions were $5 \times 5$ points on the square of $1.0~\mathrm{m} \times 1.0~\mathrm{m}$. The $\mathrm{NE}$ of Proposed was smaller than that of $\mathrm{PM}$ over the region. Note that the ITDs were accurately synthesized in both methods below $2000~\mathrm{Hz}$. 

Next, the amplitude response of the synthesized sound field was investigated. In Fig.~\ref{fig:freq_res_O}, the amplitude responses at the center of $\Omega$ of the desired sound field and synthesized sound field of Proposed and PM are plotted. Owing to the large variations in the amplitude response of PM above $2000~\mathrm{Hz}$, the timbre of the source signal can be highly distorted. In contrast, the almost flat amplitude response was achieved by Proposed at the center of $\Omega$. 

\begin{figure}[t]
    \centering
    \subfloat[PM]{\includegraphics[width=118pt,clip]{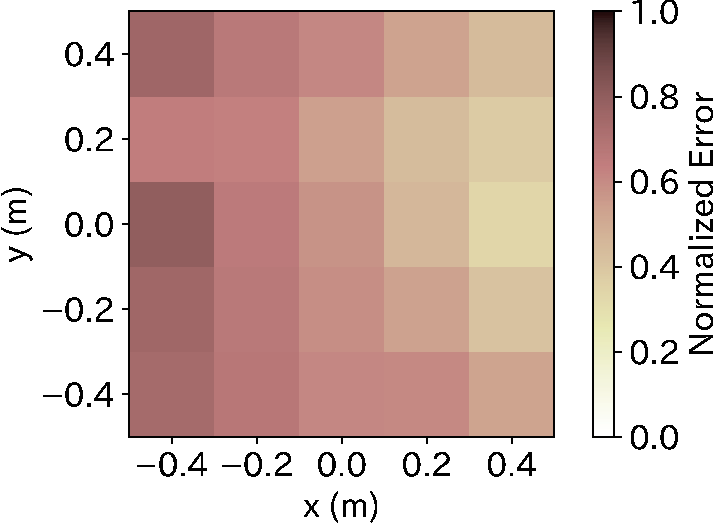}}%
    \hspace{4pt}%
    \subfloat[Proposed]{\includegraphics[width=118pt,clip]{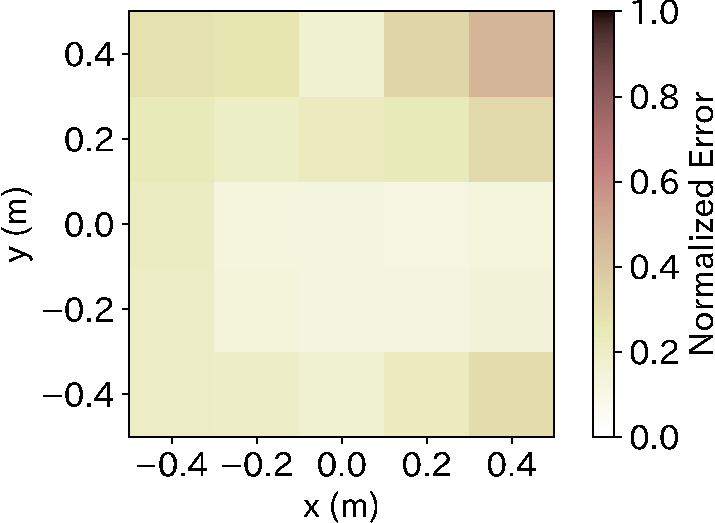}}
    \caption{Distribution of $\mathrm{NE}$ of ILDs.}
    \label{fig:ILD_NE}
\end{figure}

\begin{figure}[t]
  \centering
  \hspace*{5pt}\includegraphics[width=240pt]{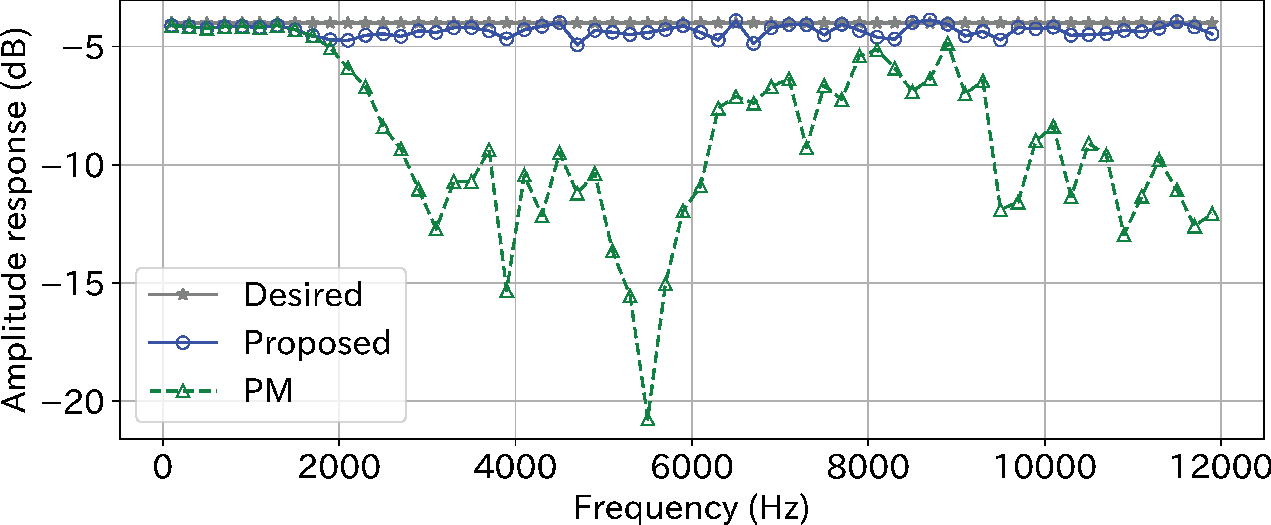}
  \caption{Amplitude responses of desired and synthesized sound fields at the center of $\Omega$.}
  \label{fig:freq_res_O}
\end{figure}

\subsection{Listening Experiments}

Listening experiments were conducted to evaluate the perceptual quality of the synthesized sound field by using a practical loudspeaker array. The numbers and positions of loudspeakers and control points were the same as those in Fig.~\ref{fig:field3D}. The reference loudspeaker was placed at $(2.0,0.5,0.0)~\mathrm{m}$ as a primary sound source of the desired field. The driving signals of Proposed and PM were obtained by assuming the loudspeakers as point sources. The reverberation time $T_{60}$ of the room was around $0.19~\mathrm{s}$. 

The perceptual quality was evaluated by using multiple stimuli with a hidden reference and anchor (MUSHRA)~\cite{MUSHRA}. Test participants were asked to rate the difference between the reference and test signals of $10~\mathrm{s}$ on a scale from 0 to 100. The reference and test signals are summarized as follows: 
\begin{itemize}
    \item Reference: The source signal played back through the reference loudspeaker.
    \item C1/Hidden anchor: The lowpass-filtered source signal up to $3.5~\mathrm{kHz}$ played back through the reference loudspeaker.
    \item C2/PM: The sound synthesized by PM and played back through the loudspeaker array.
    \item C3/Proposed: The sound synthesized by Proposed and played back through the loudspeaker array.
    \item C4/Hidden reference: The same as Reference.
\end{itemize}
The participants' head center was approximately positioned at the center of the target region by adjusting the chair, but they were able to rotate and move their heads on the chair freely. The participants were able to listen to each test signal repeatedly. Fourteen male subjects in their 20s and 30s were included, and those who scored more than 60 on the hidden anchor, which was one participant in this test, were excluded from the evaluation. Two source signals, \textbf{Vocals} and \textbf{Instrumental}, taken from track 10 of MUSDB18-HQ~\cite{MUSDB18HQ} were investigated. 

Fig.~\ref{fig:boxplot} shows the box-and-whisker plots of the scores of each test signal. The median score of Proposed was significantly higher than that of PM for both \textbf{Vocals} and \textbf{Instrumental}. After the validation of the normality of data for C2 and C3 by the Shapiro--Wilk test, Welch's $t$-test was conducted at a significance level of $0.05$. The $p$ values for \textbf{Vocals} and \textbf{Instrumental} were $9.1\times10 ^{-4}$ and $2.0\times10^{-3}$, respectively; therefore, there were significant differences in mean scores between C2 and C3 in the both cases. 

\begin{figure}[t]
    \centering
    \subfloat[\textbf{Vocals}]{\includegraphics[width=110pt]{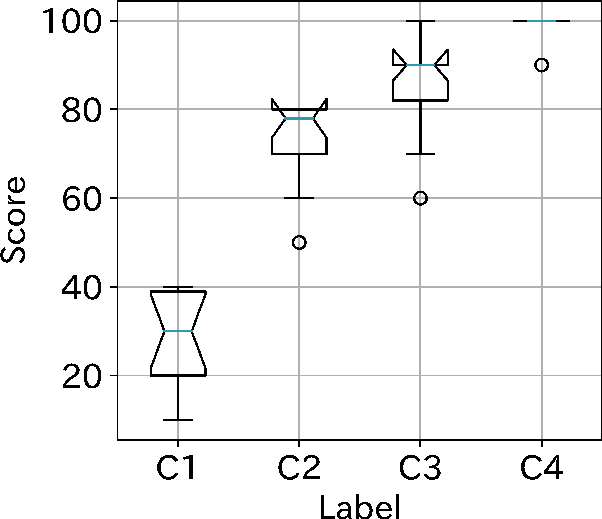}}
    \hspace{8pt}
    \subfloat[\textbf{Instrumental}]{\includegraphics[width=110pt]{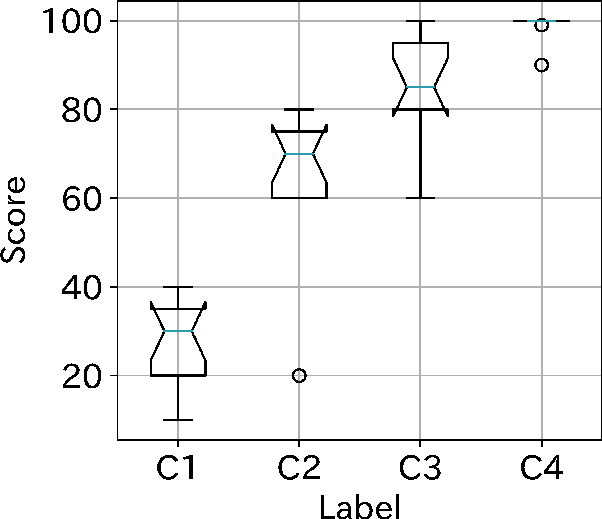}}
    \caption{Box-and-whisker plots of scores for each test signal. Circles indicate outliers, and the notches represent 95\% confidence intervals of the median.}
\label{fig:boxplot}
\end{figure}

\section{Conclusion}

We proposed a sound field synthesis method based on the combination of pressure and amplitude matching to improve perceptual quality. The cost function is defined as square errors of pressure distribution for low frequencies and amplitude distribution for high frequencies to alleviate the effects of spatial aliasing artifacts. The ADMM-based algorithm to solve this cost function is also derived. In the numerical experiments and listening experiments, it was validated that the perceptual quality of the proposed method can be improved from that of PM. Future work includes extended listening experiments to evaluate perceptual quality in more detail. 

\section{ACKNOWLEDGMENT}
\label{sec:ack}

This work was supported by JSPS KAKENHI Grant Number 22H03608 and JST FOREST Program Grant Number JPMJFR216M, Japan.

\bibliographystyle{IEEEtran}
\bibliography{str_def_abrv,koyama_en,refs}
%
%
%
%
%
%
%
%
%

\end{sloppy}
\end{document}